\documentclass[english,aps,prb, reprint,  amsmath,amssymb,floatfix,superscriptaddress,eqsecnum,longbibliography]{revtex4-2}
\usepackage[T1]{fontenc}
\usepackage[utf8]{inputenc}
\usepackage{babel}
\usepackage{amsmath}
\usepackage{amssymb}
\usepackage{graphicx}
\usepackage{xcolor}
\usepackage{braket}
\usepackage[unicode=true,pdfusetitle,
 bookmarks=true,bookmarksnumbered=false,bookmarksopen=false,
 breaklinks=false,pdfborder={0 0 1},backref=false,colorlinks=false]
 {hyperref}



\begin{document}
\title{Equation of motion and the constraining field in \textit{ab initio}
spin dynamics}
\author{Simon Streib}
\affiliation{Department of Physics and Astronomy, Uppsala University, Box 516,
SE-75120 Uppsala, Sweden}
\author{Vladislav Borisov}
\affiliation{Department of Physics and Astronomy, Uppsala University, Box 516,
SE-75120 Uppsala, Sweden}
\author{Manuel Pereiro }
\affiliation{Department of Physics and Astronomy, Uppsala University, Box 516,
SE-75120 Uppsala, Sweden}
\author{Anders Bergman}
\affiliation{Department of Physics and Astronomy, Uppsala University, Box 516,
SE-75120 Uppsala, Sweden}
\author{Erik Sjöqvist }
\affiliation{Department of Physics and Astronomy, Uppsala University, Box 516,
SE-75120 Uppsala, Sweden}
\author{Anna Delin }
\affiliation{Department of Applied Physics, School of Engineering Sciences, KTH Royal Institute of Technology, 
AlbaNova University Center, SE-10691 Stockholm, Sweden}
\affiliation{Swedish e-Science Research Center (SeRC), KTH Royal Institute of Technology, SE-10044 Stockholm, Sweden}
\author{Olle Eriksson }
\affiliation{Department of Physics and Astronomy, Uppsala University, Box 516,
SE-75120 Uppsala, Sweden}
\affiliation{School of Science and Technology, Örebro University, SE-70182 Örebro, Sweden}
\author{Danny Thonig}
\affiliation{School of Science and Technology, Örebro University, SE-70182 Örebro, Sweden}
\affiliation{Department of Physics and Astronomy, Uppsala University, Box 516,
SE-75120 Uppsala, Sweden}
\date{December 8, 2020}
\begin{abstract}
It is generally accepted that the effective magnetic field acting
on a magnetic moment is given by the gradient of the energy with respect
to the magnetization. However, in \textit{ab initio} spin dynamics within the adiabatic approximation,
the effective field is also known to be exactly the negative of the constraining
field, which acts as a Lagrange multiplier to stabilize an out-of-equilibrium, non-collinear
magnetic configuration. We show that for Hamiltonians without mean-field parameters
both of these fields are exactly equivalent, while there can be a finite
difference for mean-field Hamiltonians. For density-functional theory (DFT) calculations the constraining field obtained from the auxiliary Kohn-Sham Hamiltonian is not exactly equivalent to the DFT energy gradient. This inequality is highly relevant
for both \textit{ab initio} spin dynamics and the \textit{ab initio} calculation of
exchange constants and effective magnetic Hamiltonians. We argue that
the effective magnetic field and exchange constants have the highest accuracy in DFT when calculated
from the energy gradient and not from the constraining field.
\end{abstract}
\maketitle

\section{Introduction}

The magnetization dynamics of both insulating and metallic materials
can, in many cases, be described within the framework of atomistic spin
dynamics (see Refs.~\cite{Bertotti2009,Eriksson2017} for an overview).
This approach is valid when the electronic Hamiltonian can be mapped onto an effective model of localized spins with constant magnetic moment lengths and interaction parameters that are independent of the spin configuration. While this is generally fulfilled by magnetic insulators, these assumptions may not be
valid for magnetic metals, for magnets with non-collinear states \cite{Szilva2013,Szilva2017}, or for systems far away from equilibrium. An example of the latter is the case of ultrafast demagnetization experiments \cite{Beaurepaire1996}, where
a laser pulse demagnetizes a magnetic metal on a sub-picosecond time
scale. Such an extreme scenario would require a full non-equilibrium
treatment of the electrons, for example with time-dependent density-functional
theory (TDDFT) \cite{Runge1984}. The numerical difficulty and computational expense
of such an approach limits the applications to simulation cells with only a few atoms, which casts doubt on the ability to analyze real experiments.
Therefore, it would be desirable to have a method that combines
atomistic spin dynamics and electronic structure calculations with
a reduced numerical complexity: \textit{ab initio} spin dynamics.

Antropov \textit{et al.} proposed exactly such a formalism where the torques
on local magnetic moments are directly calculated from the electronic
ground state energy within the adiabatic approximation \cite{Antropov1995,Antropov1996}.
Stocks \textit{et al.} pointed out that an
arbitrary non-collinear magnetic configuration, which may be formed in such simulations, is not a stable ground
state of the energy functional of density functional theory (DFT), and therefore constraining fields are needed to enforce the desired magnetization
directions \cite{Stocks1998,Ujfalussy1999}. The implementation of arbitrary constraints within DFT has been worked out previously by Dederichs \textit{et al.}
\cite{Dederichs1984}. The use of constraining fields becomes essential
for magnetic configurations that deviate strongly from that of the ground state  \cite{Singer2005}.

Stocks \textit{et al.} came to the conclusion that the effective field obtained from the energy
gradient is exactly the negative of the constraining field,
as the constraining field has to cancel the effective field \cite{Stocks1998,Ujfalussy1999}. However, in this paper, the energy gradient and the constraining field are actually compared to confirm this relation. 

Here, we present such calculations for the simple case of an iron
dimer, where we find that the equivalence of the constraining field
and energy gradient is not exact. Motivated by this surprising numerical
result, we derive an exact relation between the constraining field and
energy gradient: the constraining field theorem. This theorem
shows that there is an additional term that can spoil the equality
of both fields when the Hamiltonian contains mean-field parameters,
which also applies to the auxiliary Kohn-Sham Hamiltonian in DFT calculations. We argue that the effective
field in DFT should be calculated from the energy gradient and not the constraining
field. This implies that exchange constants and effective magnetic
Hamiltonians should also be derived from the energy gradient and not from
the constraining field. 

\section{Adiabatic approximation}

The adiabatic approximation in \textit{ab initio} spin dynamics is based on
the assumption that the degrees of freedom can be separated into fast
and slow components \cite{Antropov1995,Antropov1996,Halilov1998}. The slow degree
of freedom is the magnetization direction, while the fast electronic
degrees of freedom, including the magnetic moment lengths, are assumed
to equilibrate on much shorter time scales. For the description of
the magnetization dynamics, it is then valid to consider a quasi-equilibrium
state where the magnetic moment directions are held fixed by Lagrange
multipliers that act as constraining fields on the magnetic moments.
The torques on the magnetic moments can then be calculated
from this quasi-equilibrium state.

The constrained Hamiltonian is given by
\begin{equation}
\hat{\mathcal{H}}=\hat{\mathcal{H}}_{0}+\hat{\mathcal{H}}_{\text{con}},\label{eq:Hamiltonian}
\end{equation}
where $\hat{\mathcal{H}}_{0}$ is the original Hamiltonian and the
constraining term,
\begin{equation}
\hat{\mathcal{H}}_{\text{con}}=-\sum_{i}\gamma\hat{\mathbf{S}}_{i}\cdot\mathbf{B}_{i}^{\text{con}},\label{eq:constraining term}
\end{equation}
enforces a specific quasi-equilibrium state. Here, $\hat{\mathbf{S}}_{i}$ is the operator of the total spin at site $i$, the gyromagnetic ratio $\gamma=-g|e|/(2m_{e})$ with  electron spin $g$-factor $g\approx2$, and $\mathbf{B}_{i}^{\text{con}}$
is the constraining field. 
As shown in Fig.~\ref{fig:fig1}, this field, combined with intrinsic fields that are present in $\hat{\mathcal{H}}_{0}$ in Eq.~(\ref{eq:Hamiltonian}), acts on an atomic magnetic moment, 
\begin{equation}
\mathbf{M}_{i}=\gamma\left\langle \hat{\mathbf{S}}_{i}\right\rangle =M_{i}\mathbf{m}_{i},
\end{equation}
such that the atomic moment and the constraining field are perpendicular. This causes the field to only constrain
the directions of the atomic moments, $\mathbf{m}_i$, and not the lengths, $M_{i}$.  
While Eq.~(\ref{eq:constraining term}) remains finite as an operator, its expectation value vanishes,
\begin{equation}
\left\langle \hat{\mathcal{H}}_{\text{con}}\right\rangle =-\sum_{i}\mathbf{M}_{i}\cdot\mathbf{B}_{i}^{\text{con}}=0.\label{eq:constraining energy}
\end{equation}

\section{Equation of motion\label{sec:equation of motion}}

We consider the 
equation of motion of the total spin $\hat{\mathbf{S}}_i$ at
site $i$,
\begin{equation}
\left\langle \dot{\hat{\mathbf{S}}}_{i}\right\rangle =\frac{i}{\text{\ensuremath{\hbar}}}\left\langle \left[\hat{\mathcal{H}}_{0},\hat{\mathbf{\mathbf{S}}}_{i}\right]\right\rangle ,
\end{equation}
where the expectation values have to be calculated with respect to
the ground state $\psi$ of the constrained Hamiltonian $\hat{\mathcal{H}}$. In general we have for an operator $\hat{\mathcal{O}}$,
\begin{equation}
\left\langle \hat{\mathcal{O}}\right\rangle =\bra{\psi(\{\mathbf{e}_{i}\})}\hat{\mathcal{O}}\ket{\psi(\{\mathbf{e}_{i}\})},
\end{equation}
where the ground state $\psi(\{\mathbf{e}_{i}\})$ is a function of the prescribed magnetic moment directions $\mathbf{e}_{i}$ 
with the expectation values of the moment directions fulfilling $\mathbf{m}_{i}=\mathbf{e}_{i}$.

\begin{figure}
\begin{centering}
\includegraphics[scale=2.5]{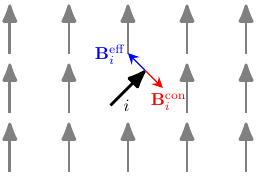}
\par\end{centering}
\caption{Effective field $\mathbf{B}_{i}^{\text{eff}}$ and constraining field $\mathbf{B}_{i}^{\text{con}}$ acting on a magnetic moment at site $i$ in a background of ferromagnetically aligned moments. \label{fig:fig1}}
\end{figure}

The
total torque on $\hat{\mathbf{S}}_i$ in the ground state of the full Hamiltonian in Eq.~(\ref{eq:Hamiltonian}) is zero. This follows from
\begin{equation}
\left\langle \left[\hat{\mathcal{H}},\hat{\mathbf{S}}_{i}\right]\right\rangle =\left\langle E_{0}\hat{\mathbf{S}}_{i}-\hat{\mathbf{S}}_{i}E_{0}\right\rangle =0,
\end{equation}
where $E_{0}$ is the ground-state energy eigenvalue of $\mathcal{H}$.
This implies that for a Hamiltonian with a constraining field one may write
\begin{equation}
\left\langle \left[\hat{\mathcal{H}}_{\text{0}},\hat{\mathbf{S}}_{i}\right]\right\rangle +\left\langle \left[\hat{\mathcal{H}}_{\text{con}},\hat{\mathbf{S}}_{i}\right]\right\rangle =0.\label{eq:constrained eom}
\end{equation}
We identify
\begin{align}
\left\langle \dot{\hat{\mathbf{S}}}_{i}\right\rangle  & =\frac{i}{\text{\ensuremath{\hbar}}}\left\langle \left[\hat{\mathcal{H}}_{\text{0}},\hat{\mathbf{S}}_{i}\right]\right\rangle \nonumber \\
 & =\gamma\left\langle \hat{\mathbf{S}}_{i}\right\rangle \times\mathbf{B}_{i}^{\text{eff}},\label{eq:constrained eom2}
\end{align}
where $\mathbf{B}_{i}^{\text{eff}}$ is the effective field that drives the dynamics of ${\hat{\mathbf{S}}}_{i}$. Only the component of $\mathbf{B}_{i}^{\text{eff}}$ perpendicular to ${\mathbf{S}}_{i}$ contributes to the equation of motion and we define the parallel component of $\mathbf{B}_{i}^{\text{eff}}$ to be zero. By combining Eqs.~(\ref{eq:constrained eom}) and (\ref{eq:constrained eom2}), one obtains
\begin{align}
\left\langle \dot{\hat{\mathbf{S}}}_{i}\right\rangle  & =-\frac{i}{\text{\ensuremath{\hbar}}}\left\langle \left[\hat{\mathcal{H}}_{\text{con}},\hat{\mathbf{S}}_{i}\right]\right\rangle \nonumber \\
 & =\gamma\left\langle \hat{\mathbf{S}}_{i}\right\rangle \times\left(-\mathbf{B}_{i}^{\text{con}}\right),\label{eq:torque}
\end{align}
which implies that the effective field
\begin{equation}
\mathbf{B}_{i}^{\text{eff}}=-\mathbf{B}_{i}^{\text{con}}. \label{eq:effective field}
\end{equation}
The constraining field cancels the effective field, as illustrated in Fig.~\ref{fig:fig1}.
This shows that the correct torque for a given Hamiltonian $\hat{\mathcal{H}}_{0}$
can be obtained from the constraining field. 

For a classical spin system with localized rigid spins (e.g., the Heisenberg model discussed in Appendix\,\ref{sec:Spin Hamiltonians}) it can be shown that
the effective field is exactly \cite{Landau35,Keshtgar2017} 
\begin{equation}
\mathbf{B}_{i}^{\text{cl}}=-\boldsymbol{\nabla}_{\mathbf{M}_{i}}\mathcal{\mathcal{H}}_{0},\label{eq:classical effective field}
\end{equation}
which would offer an alternative approach of calculating the effective field of atomistic spin-dynamics, $\mathbf{B}_{i}^{\text{eff}}$, compared to Eq.~(\ref{eq:effective field}).
However, it is not obvious that this should also hold for an itinerant magnet
where the Hamiltonian is not a simple function of
spin operators, but of the creation and annihilation operators of the itinerant electrons. Below we evaluate the effective field from the two approaches (\ref{eq:effective field}) and (\ref{eq:classical effective field}).

\section{Constraining field}

In the previous section we introduced the constraining field $\mathbf{B}_{i}^{\text{con}}$,
but we have so far not given a procedure how to calculate this field.
We know that the constraining field at site $i$ has to be tuned so that it cancels the intrinsic field of the Hamiltonian $\hat{\mathcal{H}}_{0}$ and is perpendicular to the magnetic moment direction $\mathbf{m}_{i}$. 
Furthermore, the constraining field has to be chosen such that at each site the moment points along
the prescribed moment direction $\mathbf{e}_{i}$,
\begin{equation}
\mathbf{m}_{i}\overset{!}{=}\mathbf{e}_{i}.
\end{equation}
We are going to consider here two methods of calculating the constraining
field: the method proposed by Stocks \textit{et al.} \cite{Stocks1998,Ujfalussy1999}
and the method by Ma and Dudarev \cite{Ma2015}.

Stocks \textit{et al.} provide the following iterative procedure for calculating
the constraining field \cite{Stocks1998,Ujfalussy1999},
\begin{align}
\mathbf{B}_{i}^{\text{con}}(k+1) & =\mathbf{B}_{i}^{\text{con}}(k)-\left(\mathbf{B}_{i}^{\text{con}}(k)\cdot\mathbf{e}_{i}\right)\mathbf{e}_{i}\nonumber \\
 & -B_{0}\left[\mathbf{m}_{i}-\left(\mathbf{m}_{i}\cdot\mathbf{e}_{i}\right)\mathbf{e}_{i}\right],\label{eq:Stocks}
\end{align}
where $k$ is the iteration index and $B_{0}$ is a free parameter that
can be tuned for optimal convergence. The first two terms of Eq.~(\ref{eq:Stocks})
ensure that only the contribution perpendicular to $\mathbf{e}_{i}$
is carried to the next iteration, while the third term adjusts the
constraining field by a term proportional to the difference between
the output and prescribed moment direction (again only keeping the
perpendicular contribution), which aligns the magnetic moment $\mathbf{m}_{i}$
closer along the prescribed direction $\mathbf{e}_{i}$. The algorithm given by Eq.~(\ref{eq:Stocks}) can be derived systematically from the method
of Lagrange multipliers \cite{Ivanov2020}. The uniqueness of the constraining field follows from the uniqueness of the solutions within
constrained DFT \cite{Wu2005}.

Ma and Dudarev derive the constraining field by imposing an energy
penalty for misalignments of $\mathbf{m}_{i}$ and $\mathbf{e}_{i}$,
which leads to the constraining field \cite{Ma2015}
\begin{equation}
\mathbf{B}_{i}^{\text{con}}=-2\lambda\left[\mathbf{m}_{i}-\left(\mathbf{m}_{i}\cdot\mathbf{e}_{i}\right)\mathbf{e}_{i}\right],\label{eq:Ma}
\end{equation}
where $\lambda$ determines the strength of the energy penalty and
convergence is formally reached for $\lambda\to\infty$ with $\mathbf{m}_{i}\approx\mathbf{e}_{i}+\mathcal{O}(\lambda^{-1})$
\cite{Ma2015}. 

Both methods look similar, but have different advantages and disadvantages.
The first method (\ref{eq:Stocks}) has the advantage of good convergence
since the constraining field is adjusted step by step, but requires
this additional iterative calculation, which can be done in parallel to a self-consistent calculation. The second method (\ref{eq:Ma})
has the advantage that it does not introduce an additional iterative
calculation and can be simply included in a self-consistent calculation,
but it has the disadvantage that convergence is problematic
if $\lambda$ is too large. This convergence problem can be circumvented by increasing the value of $\lambda$ in steps, which keeps the energy penalty sufficiently small.

Within the formalism of constrained DFT, the constraining field can be directly included as a Lagrange multiplier
in the energy minimization procedure to determine the constrained ground state \cite{Dederichs1984,Kurz2004,Singer2005,Cuadrado2018,Ivanov2020}.

\section{Numerical results for a dimer}

\begin{figure}
\begin{centering}
\includegraphics[width=1.0\columnwidth]{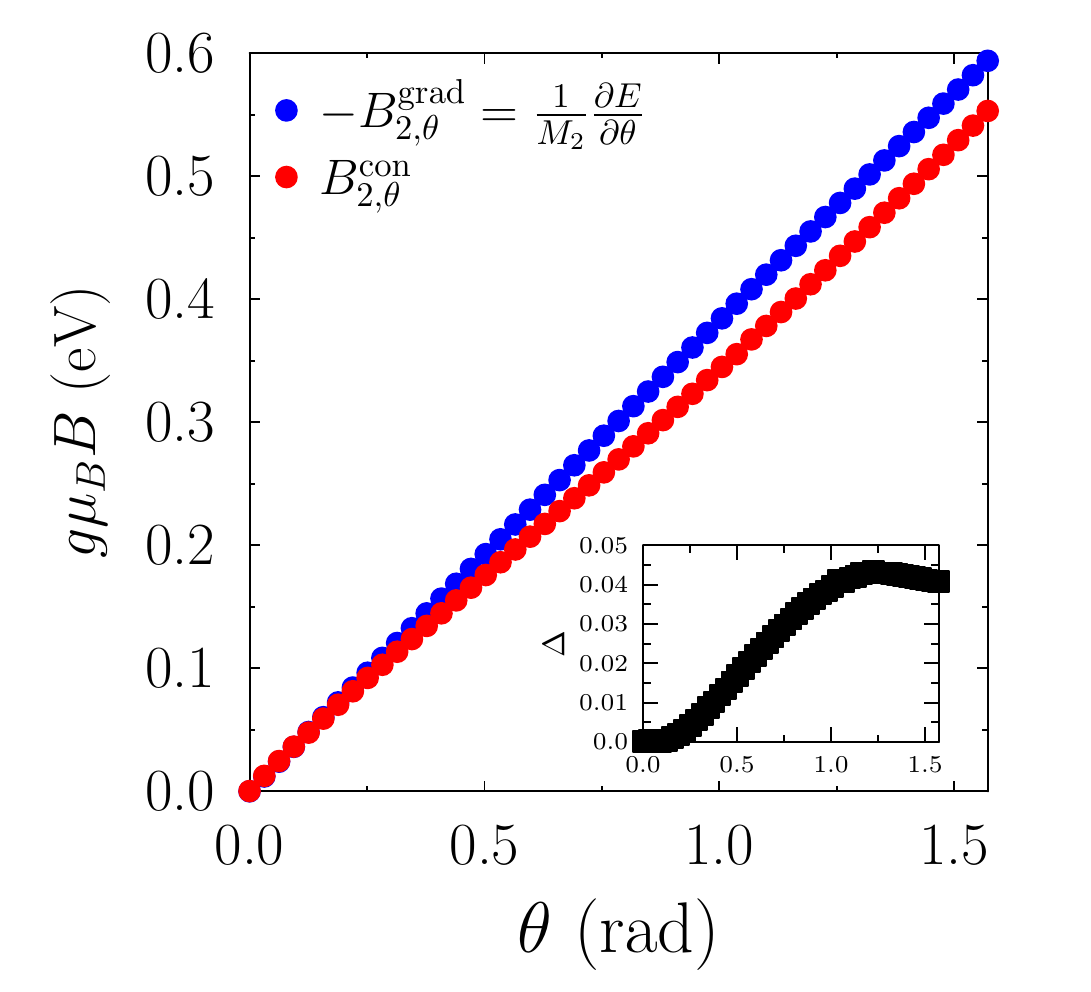}
\par\end{centering}
\caption{Comparison of the effective magnetic field calculated from the constraining
field $B^{\mathrm{con}}_{2,\theta}$ (red symbols) and the energy gradient $B_{2,\theta}^{\mathrm{grad}}$ (blue symbols) for an iron dimer, where the
moment $i=2$ is rotated by $\theta$. The inset shows the difference, $\Delta=-B^{\mathrm{grad}}_{2,\theta}-B^{\mathrm{con}}_{2,\theta}$, between the two effective fields. \label{fig:VASP}}
\end{figure}

To investigate the relation between the constraining field and the
effective field obtained from the energy gradient, we performed DFT
calculations for an Fe dimer system with VASP \cite{Kresse1996,Kresse1996b}.
The constraining field implementation in VASP is based on the method
by Ma and Dudarev \cite{Ma2015}, as described above. See Appendix \ref{sec:Numerical_Details} for more details.

We start with both magnetic moments of the two iron atoms aligned
along the $z$ axis and rotate one of the magnetic moments by an angle
$\theta$ in the $xz$ plane, which gives an expression for the moment of the second atom,
\begin{align}
m_{2}^{x} & =\sin\theta,\;m_{2}^{y}=0,\;m_{2}^{z}=\cos\theta.
\end{align}
The effective field calculated from the energy gradient is given by
\begin{equation}
\mathbf{B}_{i}^{\text{grad}}=-\frac{1}{M_{i}}\boldsymbol{\nabla}_{\mathbf{e}_{i}}\left\langle \hat{\mathcal{H}}_{0}\right\rangle.
\label{gradfield}
\end{equation}
Since $\mathbf{e}_i$ is a unit vector with a fixed length, the gradient has to be defined as
\begin{equation}
    \boldsymbol{\nabla}_{\mathbf{e}_{i}}f=\frac{\partial f}{\partial \theta_i} \hat{\boldsymbol{\theta}}_i + \frac{1}{\sin\theta_i} \frac{\partial f}{\partial \phi_i}\hat{\boldsymbol{\phi}}_i,\label{eq:gradient}
\end{equation}
where $\theta_i$ and $\phi_i$ are the polar and azimuthal angles in spherical coordinates with their corresponding unit vectors $\hat{\boldsymbol{\theta}}_i$ and $\hat{\boldsymbol{\phi}}_i$.
The $\theta$ component of Eq.~(\ref{gradfield}) is therefore
\begin{equation}
{B}_{i,\theta}^{\text{grad}}=-\frac{1}{M_i} \frac{\partial}{\partial \theta_i}\left\langle \hat{\mathcal{H}}_{0}\right\rangle.    
\end{equation}

In Fig.~\ref{fig:VASP}, we show the $\theta$ component of the constraining field acting on the rotated spin and we compare this field to what one obtains from the energy gradient. The two calculations, Eqs.~(\ref{eq:effective field}) and (\ref{gradfield}), are plotted in Fig.~\ref{fig:VASP} as a function of $\theta$. From the figure one concludes that the two
fields are similar, but not exactly identical. It is also possible to discern that the difference between them becomes bigger the further away one is from the equilibrium configuration.

\begin{figure}
\begin{centering}
\includegraphics[scale=1]{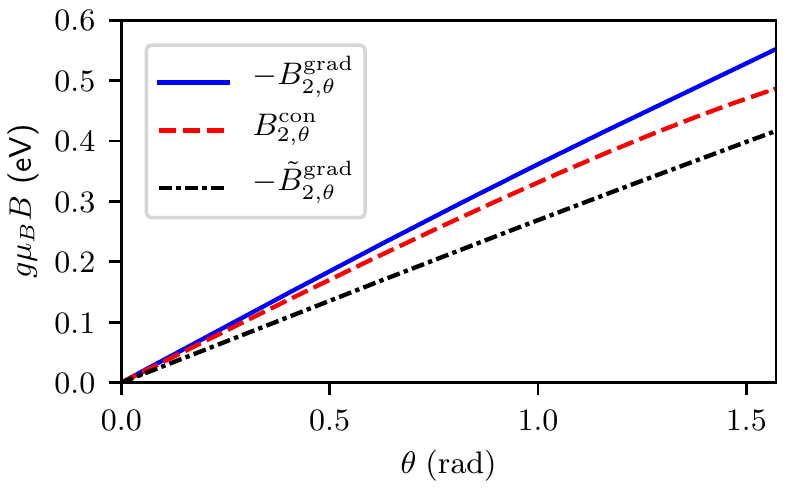}
\par\end{centering}
\caption{Comparison of the effective magnetic field calculated from the constraining
field and the energy gradient for our mean-field tight-binding model applied to an iron dimer where the moment $i=2$ is rotated by $\theta$. The field $\tilde{B}_{2,\theta}^{\text{grad}}$ is calculated without constraining fields (see Appendix \ref{sec:Tight-binding}).\label{fig:TB fields}}
\end{figure}

For comparison, we performed analogous calculations with a mean-field tight-binding model (see Appendix~\ref{sec:Tight-binding} for details), which, as can be seen in Fig.~\ref{fig:TB fields}, show similar results. There we also show the field $\tilde{\mathbf{B}}_{i}^{\text{grad}}$ which is calculated without constraining fields, with the constraints only implemented approximately by imposing local quantization axes \cite{Grotheer2001} (see Appendix \ref{sec:Tight-binding}). This approximate method underestimates in our case the gradient field by about 25\%, even in the limit $\theta \to 0$. This implies an underestimate by 25\% of the exchange parameter $J$ of the dimer in the absence of constraining fields, see Appendix \ref{sec:Spin Hamiltonians}. The widely used Liechtenstein-Katsnelson-Antropov-Gubanov (LKAG) formalism \cite{Liechtenstein1984,Liechtenstein1987} for the calculation of exchange parameters does not take constraining fields into account \cite{Bruno2003}, which could potentially result in similar inaccuracies \cite{Jacobsson2017}.

In the following, we analyze
the origin of the difference between the constraining and energy gradient fields and we argue why it matters
to be aware of this difference.

\section{Constraining field theorem}

In this section, we derive the constraining field theorem, the main result of this paper, which relates the constraining field and the energy gradient field. We discuss under which circumstances these fields are equivalent and the implications this has for the correct choice of the effective field in spin-dynamics simulations.

\subsection{Derivation of the theorem}

To relate the energy gradient field
$\mathbf{B}_{i}^{\text{grad}}$
to the constraining field $\mathbf{B}_{i}^{\text{con}}$, we wish to evaluate $\boldsymbol{\nabla}_{\mathbf{e}_{i}}\langle \hat{\mathcal{H}}_{0}+\hat{\mathcal{H}}_{\text{con}}\rangle$. Here, and in the following, an expression 
\begin{equation}
   \boldsymbol{\nabla}_{\mathbf{e}_{i}}\left\langle \hat{\mathcal{O}} \right\rangle = \left\langle \boldsymbol{\nabla}_{\mathbf{e}_{i}}\hat{\mathcal{O}}\right\rangle + \boldsymbol{\nabla}_{\mathbf{e}_{i}}^{\psi}\left\langle \hat{\mathcal{O}} \right\rangle 
\end{equation}
is used, where the first term on the right-hand side accounts for the gradient of the operator $\hat{\mathcal{O}}$ and the second accounts for the gradient (or rotation) of the wavefunction $\psi$ used to calculate the expectation value. We obtain the Hellmann-Feynman theorem \cite{Hellmann1937,Feynman1939},
\begin{equation}
\boldsymbol{\nabla}_{\mathbf{e}_{i}}\left\langle \hat{\mathcal{H}}_{0}+\hat{\mathcal{H}}_{\text{con}}\right\rangle =\left\langle \boldsymbol{\nabla}_{\mathbf{e}_{i}}\hat{\mathcal{H}}_{0}+\boldsymbol{\nabla}_{\mathbf{e}_{i}}\hat{\mathcal{H}}_{\text{con}}\right\rangle ,\label{eq:Hellmann-Feynman}
\end{equation}
where the term that accounts for the wavefunction variation vanishes due to the extremity of the ground-state energy,
\begin{equation}
    \boldsymbol{\nabla}_{\mathbf{e}_{i}}^{\psi}\left\langle \hat{\mathcal{H}}_{\text{0}}+\hat{\mathcal{H}}_{\text{con}}\right\rangle = 0. \label{eq:variational principle}
\end{equation}
The derivative of the constraining part can be expressed as
\begin{equation}
\boldsymbol{\nabla}_{\mathbf{e}_{i}}\left\langle \hat{\mathcal{H}}_{\text{con}}\right\rangle =\left\langle \boldsymbol{\nabla}_{\mathbf{e}_{i}}\hat{\mathcal{H}}_{\text{con}}\right\rangle +\boldsymbol{\nabla}_{\mathbf{e}_{i}}^{\psi}\left\langle \hat{\mathcal{H}}_{\text{con}}\right\rangle.
\end{equation}
Here one should note that the eigenstate that minimizes the expression in Eq.~(\ref{eq:variational principle}), does not necessarily imply that a variation over the wavefunction vanishes when one considers only $\hat{\mathcal{H}}_{\text{con}}$, i.e., $\boldsymbol{\nabla}_{\mathbf{e}_{i}}^{\psi}\langle \hat{\mathcal{H}}_{\text{con}}\rangle$ is non-zero. 
Similarly, both terms need to be considered when treating only $\hat{\mathcal{H}}_{0}$ in the variation,
\begin{equation}
\boldsymbol{\nabla}_{\mathbf{e}_{i}}\left\langle \hat{\mathcal{H}}_{0}\right\rangle =\boldsymbol{\nabla}_{\mathbf{e}_{i}}^{\psi}\left\langle \hat{\mathcal{H}}_{0}\right\rangle +\left\langle \boldsymbol{\nabla}_{\mathbf{e}_{i}}\hat{\mathcal{H}}_{0}\right\rangle.
\end{equation}
Using the relationship in Eq.~(\ref{eq:variational principle}) leads to
\begin{equation}
\boldsymbol{\nabla}_{\mathbf{e}_{i}}\left\langle \hat{\mathcal{H}}_{0}\right\rangle =-\boldsymbol{\nabla}_{\mathbf{e}_{i}}^{\psi}\left\langle \hat{\mathcal{H}}_{\text{con}}\right\rangle +\left\langle \boldsymbol{\nabla}_{\mathbf{e}_{i}}\hat{\mathcal{H}}_{0}\right\rangle.
\label{con1}
\end{equation}
Since the moment directions for $j\neq i$ are held constant and $\mathbf{M}_{i}\cdot\mathbf{B}_{i}^{\text{con}}=0$,
we find
\begin{equation}
-\frac{1}{M_{i}}\boldsymbol{\nabla}_{\mathbf{e}_{i}}^{\psi}\left\langle \hat{\mathcal{H}}_{\text{con}}\right\rangle   =\sum_{\alpha=x,y,z}B_{i,\alpha}^{\text{con}}\boldsymbol{\nabla}_{\mathbf{e}_{i}}^{\psi}m_{i}^{\alpha}=\mathbf{B}_{i}^{\text{con}}.\label{con2}
\end{equation}
From Eqs.~(\ref{gradfield}), (\ref{con1}), and (\ref{con2}), we arrive at our main result, the constraining field theorem:
\begin{equation}
\mathbf{B}_{i}^{\text{grad}}=-\mathbf{B}_{i}^{\text{con}}-\frac{1}{M_{i}}\left\langle \boldsymbol{\nabla}_{\mathbf{e}_{i}}\hat{\mathcal{H}}_{0}\right\rangle .\label{eq:constraining field theorem}
\end{equation}

For a Hamiltonian without any mean-field parameters there is no  dependence on the moment directions, $\boldsymbol{\nabla}_{\mathbf{e}_{i}}\hat{\mathcal{H}}_{0}=0$,
which directly implies
\begin{equation}
\mathbf{B}_{i}^{\text{eff}}=-\mathbf{B}_{i}^{\text{con}}=\mathbf{B}_{i}^{\text{grad}}.
\end{equation}
This is analogous to the case of an effective spin Hamiltonian, given by
Eq.~(\ref{eq:classical effective field}), where the effective field is also given
by the energy gradient.

\begin{figure}
\begin{centering}
\includegraphics[scale=1]{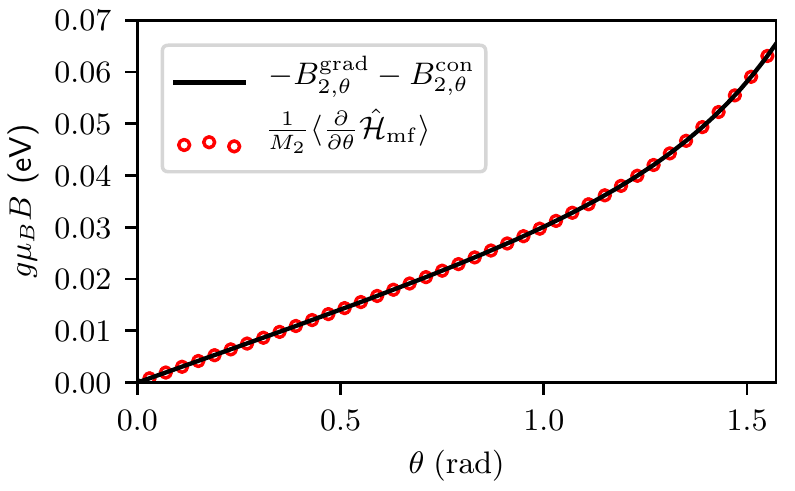}
\par\end{centering}
\caption{Numerical check of the constraining field theorem, Eq.~(\ref{eq:constraining field theorem}), for a mean-field tight-binding model applied to an iron dimer where the moment $i=2$ is rotated by $\theta$. \label{fig:numerical check}}
\end{figure}

\subsection{Mean-field Hamiltonians}

In a mean-field treatment, the Hamiltonian $\hat{\mathcal{H}}_{\text{0}}$ is symbolized with $\hat{\mathcal{H}}_{\text{mf}}$, where the parameters of the Hamiltonian in general depend on the directions $\{\mathbf{e}_{i}\}$. In this case, the
term $\langle\boldsymbol{\nabla}_{\mathbf{e}_{i}}\hat{\mathcal{H}}_{\text{mf}}\rangle$
can be finite, and it follows from Eq.~(\ref{eq:constraining field theorem})
that the relation $\mathbf{B}_{i}^{\text{grad}}=-\mathbf{B}_{i}^{\text{con}}$
is not exact. Figure~\ref{fig:numerical check} shows that this term gives precisely the difference between constraining and energy gradient fields for a mean-field tight-binding calculation (see  Appendix \ref{sec:Tight-binding} for details), supporting the constraining field theorem (\ref{eq:constraining field theorem}). 
The difference between the two fields is determined by how strongly the mean-field parameters depend on the moment directions $\{\mathbf{e}_{i}\}$ and by how strong the correlation effects are that are represented by the mean-field contribution to $\hat{\mathcal{H}}_{\text{mf}}$. If $\hat{\mathcal{H}}_{\text{mf}}$ includes the operator-independent energy contributions arising from the mean-field decoupling, this leads to a cancellation such that $\langle\boldsymbol{\nabla}_{\mathbf{e}_{i}}\hat{\mathcal{H}}_{\text{mf}}\rangle=0$ and $\mathbf{B}_{i}^{\text{grad}}=-\mathbf{B}_{i}^{\text{con}}$, as we demonstrate in Appendix\,\ref{sec:Mf decoupling}. Such operator-independent energy contributions are not included in the tight-binding calculations shown in Figs.~\ref{fig:TB fields} and \ref{fig:numerical check}.

\subsection{Density functional theory}

For the DFT calculations, we have to consider the auxiliary Kohn-Sham (KS) Hamiltonian \cite{Kohn1965} (here without spin-orbit interaction),
\begin{equation}
\hat{\mathcal{H}}_{\mathrm{KS}}=\sum_l\left[\frac{\hat{\mathbf{p}}_l^2}{2m_e}+V_{\mathrm{eff}}(\hat{\mathbf{r}}_l,\hat{\mathbf{S}}_l)\right],
\end{equation}
where $l$ is the index of the KS quasiparticle with position and momentum operators $\hat{\mathbf{r}}_l$ and $\hat{\mathbf{p}}_l$ and spin operator $\hat{\mathbf{S}}_l$. The effective potential $V_{\mathrm{eff}}$ 
is not only dependent on the position and spin of the quasiparticle, but also 
depends on the average electron and magnetization densities, $n(\mathbf{r})$ and $\mathbf{M}(\mathbf{r})$, and we write
\begin{align}
V_{\mathrm{eff}}(\hat{\mathbf{r}}_l,\hat{\mathbf{S}}_l)=&V_{\mathrm{ext}}(\hat{\mathbf{r}}_l)+\int \frac{n(\mathbf{r}')e^2}{4\pi\varepsilon_0|\hat{\mathbf{r}}_l-\mathbf{r}'|}
\,\mathrm{d}^3 r' + \mu_{\mathrm{xc}}(\hat{\mathbf{r}}_l)\nonumber\\
& -\gamma\hat{\mathbf{S}}_l\cdot \left[ \mathbf{B}_{\mathrm{ext}}(\hat{\mathbf{r}}_l)+\mathbf{B}_{\mathrm{xc}}(\hat{\mathbf{r}}_l)\right],
\end{align}
where $V_{\mathrm{ext}}$ is the Coulomb potential from the ions in the lattice and $\mathbf{B}_\mathrm{ext}$ is an external magnetic field. The scalar and magnetic exchange-correlation potentials are given by
\begin{align}
\mu_{\mathrm{xc}}(\mathbf{r})=&\frac{\delta E_{\mathrm{xc}}[n,\mathbf{M}]}{\delta n(\mathbf{r})},\\
\mathbf{B}_{\mathrm{xc}}(\mathbf{r})=&-\frac{\delta E_{\mathrm{xc}}[n,\mathbf{M}]}{\delta \mathbf{M}(\mathbf{r})},
\end{align}
where $E_{\mathrm{xc}}$ is the exchange-correlation energy, which is a functional of the electron and magnetization densities. Since these densities depend on the magnetic moment directions $\{\mathbf{e}_{i}\}$, we have $\boldsymbol{\nabla}_{\mathbf{e}_{i}}\hat{\mathcal{H}}_{\mathrm{KS}}\neq0$ and therefore $\mathbf{B}_{i}^{\text{con}}\neq-\mathbf{B}_{i,\mathrm{KS}}^{\text{grad}}$, according to Eq.~(\ref{eq:constraining field theorem}) applied to $\hat{\mathcal{H}}_\mathrm{KS}$. 

It is important to note here that the constraining field theorem (\ref{eq:constraining field theorem}) is applied to the KS Hamiltonian and the energy gradient field $\mathbf{B}_{i,\mathrm{KS}}^{\text{grad}}$ is therefore given by the gradient of the KS energy $\langle \hat{\mathcal{H}}_\mathrm{KS}\rangle$ and not of the total DFT energy $E_\mathrm{DFT}$, which contains an additional double counting term $E_\mathrm{dc}$ \cite{Kohn1965,Liechtenstein1987},
\begin{equation}
    {E}_{\mathrm{DFT}}=\left\langle \hat{\mathcal{H}}_{\mathrm{KS}}\right\rangle + E_\mathrm{dc}.
\end{equation}
This implies for the energy gradient of the total DFT energy with Eq.~(\ref{eq:constraining field theorem}) applied to the KS Hamiltonian,
\begin{align}
    \mathbf{B}_{i,\mathrm{DFT}}^{\text{grad}}&=-\mathbf{B}_{i}^{\text{con}}-\frac{1}{M_{i}}\left\langle \boldsymbol{\nabla}_{\mathbf{e}_{i}}\hat{\mathcal{H}}_{\mathrm{KS}}\right\rangle -\frac{1}{M_i}\boldsymbol{\nabla}_{\mathbf{e}_{i}}E_\mathrm{dc}\nonumber \\
    &=-\mathbf{B}_{i}^{\text{con}}-\frac{1}{M_{i}}\left\langle \boldsymbol{\nabla}^*_{\mathbf{e}_{i}}\hat{\mathcal{H}}_{\mathrm{KS}}\right\rangle,\label{DFT theorem}
\end{align}
where the last two terms of the first line cancel only partially \footnote{The result  $\langle\boldsymbol{\nabla}_{\mathbf{e}_{i}}\hat{\mathcal{H}}_{\mathrm{KS}}\rangle+\boldsymbol{\nabla}_{\mathbf{e}_{i}}E_\mathrm{dc}=\langle\boldsymbol{\nabla}^*_{\mathbf{e}_{i}}\hat{\mathcal{H}}_{\mathrm{KS}}\rangle$ follows from the derivation in Appendix A of Ref.~\cite{Liechtenstein1987}.} and $\boldsymbol{\nabla}^*_{\mathbf{e}_{i}}$ denotes the variation at fixed $n(\mathbf{r})$ and $M(\mathbf{r})=|\mathbf{M}(\mathbf{r})|$. Equation (\ref{DFT theorem}) is the DFT adaptation of the constraining field theorem (\ref{eq:constraining field theorem}) and explains why the DFT calculations in Fig.~\ref{fig:VASP} show a difference between the constraining and energy gradient fields.

This difference depends on the non-collinearity of the magnetization density and vanishes near the collinear limit (see Fig.\,\ref{fig:VASP}). We can confirm this by considering that in this case the exchange-correlation field is approximately collinear within the volume $\Omega_i$ that is associated with the atomic site $i$,
\begin{equation}
\mathbf{B}_\mathrm{xc}(\mathbf{r})\sim \mathbf{e}_i, \quad\forall \mathbf{r}\, \in \Omega_i.
\end{equation}
We find
\begin{align}
    \left\langle \boldsymbol{\nabla}^*_{\mathbf{e}_{i}}\hat{\mathcal{H}}_{\mathrm{KS}}\right\rangle&=-\int \left[\boldsymbol{\nabla}^*_{\mathbf{e}_{i}}\mathbf{B}_\mathrm{xc}(\mathbf{r})\right]\cdot \mathbf{M}(\mathbf{r}) \,\mathrm{d}^3 r \nonumber\\
    &\sim \int_{\Omega_i} \mathbf{M}(\mathbf{r})\,\mathrm{d}^3 r=\mathbf{M}_i,
\end{align}
which does not contribute to the effective field since only components perpendicular to $\mathbf{M}_i$ contribute. For bulk systemsm we therefore expect that the difference between constraining and energy gradient fields will be more pronounced for short-wavelength spin waves due to their stronger non-collinearity.

\section{Choice of the effective field}

For a Hamiltonian $\hat{\mathcal{H}}_{0}$ with $\langle\boldsymbol{\nabla}_{\mathbf{e}_{i}}\hat{\mathcal{H}}_0\rangle=0$ it does not matter
if the effective field is calculated from the energy gradient or the
constraining field. But what is the right choice for the effective
field if that is not the case, and in particular, what choice should one make in calculations based on DFT?

Let us first assume that the DFT energy $E_{\text{DFT}}$
exactly reproduces the energies of the Hamiltonian $\hat{\mathcal{H}}_{0}$,
i.e,. for each configuration $\{\mathbf{e}_{i}\}$,
\begin{equation}
E_{\text{DFT}} =\left\langle \hat{\mathcal{H}}_{0}\right\rangle .\label{eq:energy equality}
\end{equation}
The correct effective field is then given by the DFT energy gradient,
\begin{equation}
\mathbf{B}_{i}^{\text{eff}}=-\frac{1}{M_{i}}\boldsymbol{\nabla}_{\mathbf{e}_{i}}\left\langle \hat{\mathcal{H}}_{0}\right\rangle =-\frac{1}{M_{i}}\boldsymbol{\nabla}_{\mathbf{e}_{i}}{E}_{\text{DFT}}=\mathbf{B}_{i,\mathrm{DFT}}^{\text{grad}},
\end{equation}
which is not exactly the same as the negative of the constraining field obtained from the KS Hamiltonian, as shown by Eq.~(\ref{DFT theorem}) and Fig.~\ref{fig:VASP}. This implies that the constraining field of the KS Hamiltonian $\hat{\mathcal{H}}_\mathrm{KS}$ is not the same as the one of the original Hamiltonian $\hat{\mathcal{H}}_0$ and therefore $\hat{\mathcal{H}}_\mathrm{KS}$ does not exactly reproduce the equation of motion,
\begin{equation}
\left\langle \left[\hat{\mathcal{H}}_{\mathrm{KS}},\hat{\mathbf{\mathbf{S}}}_{i}\right]\right\rangle\neq \left\langle \left[\hat{\mathcal{H}}_{0},\hat{\mathbf{\mathbf{S}}}_{i}\right]\right\rangle.
\end{equation}
This is not a failure of DFT since the DFT formalism is designed to provide the correct ground state energies and electron densities. The KS Hamiltonian cannot be used to correctly describe non-equilibrium physics.

When the constraining field is not equivalent to the energy gradient, it is not exact to construct an effective magnetic Hamiltonian based on the calculation of the constraining field alone. The exchange parameters
have to be calculated from the energy gradient \cite{Liechtenstein1984,Liechtenstein1987,Bruno2003}.

If we are not considering DFT calculations and we cannot make the assumption (\ref{eq:energy equality}), then the argument above does not apply. The effective field describing the magnetization dynamics of a given Hamiltonian $\hat{\mathcal{H}}_0$ is then the negative of the constraining field, as shown in Sec.~\ref{sec:equation of motion}.

\section{Summary}

We have shown that the effective magnetic field in the equation of
motion within the adiabatic approximation is exactly the negative of the constraining
field. For Hamiltonians that do not contain mean-field parameters
depending on the moment directions, the effective field derived from
the energy gradient is equivalent to the constraining field. We have argued that in the case of DFT the effective field should
be calculated from the energy gradient because DFT is designed to reproduce the
physically correct energies. 

Our results have three important implications: 

(1) In \textit{ab initio} spin dynamics, the constraining field alone may be insufficient for calculations of 
the effective field, which should be obtained from the energy gradient.

(2) Therefore, exchange constants for an effective magnetic Hamiltonian also should
be calculated from energy gradients and not from the constraining
fields. 

(3) Our tight-binding calculations support the notion that an approximate implementation of out-of-equilibrium, non-collinear states without constraining fields can give inaccurate results, even in the vicinity of the ferromagnetic ground state.

\begin{acknowledgments}
We thank Pavel Bessarab, Mikhail Katsnelson, Alexander Lichtenstein, and Attila Szilva for helpful discussions. A.B. acknowledges discussions with Pui-Wai Ma.
The authors acknowledge financial support from the Knut and Alice Wallenberg Foundation through Grant No. 2018.0060. O.E. also acknowledges support of eSSENCE, the Swedish Research Council (VR), the Foundation for Strategic Research (SSF) and the ERC (synergy grant).  D.T. acknowledges support from the Swedish Research Council (VR)  through  Grant  No.   2019-03666. A.D. acknowledges support from the Swedish Research Council (VR). 
The computations were enabled by resources provided by the
Swedish National Infrastructure for Computing (SNIC) at Chalmers Center for Computational Science and Engineering (C3SE), High Performance Computing Center North (HPCN), and the National Supercomputer Center (NSC) partially funded by the Swedish Research Council through Grant Agreement No. 2016-07213.
\end{acknowledgments}

\appendix

\section{Tight-binding model\label{sec:Tight-binding}}

The tight-binding model considered here is given by the Hamiltonian
\begin{equation}
\hat{\mathcal{H}}_{\text{tb}}=\sum_{i\alpha,j\beta,\sigma}t_{i\alpha,j\beta}\hat{c}_{i\alpha\sigma}^{\dagger}\hat{c}_{j\beta\sigma}+\hat{\mathcal{H}}_{\text{Stoner}}+\hat{\mathcal{H}}_{\text{lcn}},
\end{equation}
where $\hat{c}_{i\alpha\sigma}^{\dagger}$ and $\hat{c}_{i\alpha\sigma}$ are
the creation and annihilation operators of electrons at site $i$
in the orbital state $\alpha$ with spin $\sigma$. The matrix elements
$t_{i\alpha,j\beta}$ are based on a Slater-Koster parametrization \cite{Slater1954} with the parameters taken from Ref.~\cite{Thonig2014}. The Stoner term is defined
as
\begin{equation}
\hat{\mathcal{H}}_{\text{Stoner}}=\sum_{i,\alpha}\frac{I_{\alpha}}{\hbar\mu_B}M_{i,\alpha}\mathbf{e}_{i}\cdot\hat{\mathbf{S}}_{i,\alpha},\label{eq:Stoner}
\end{equation}
where $I_{\alpha}$ is the Stoner parameter for orbital $\alpha$,
$M_{i,\alpha}$ is the moment length associated with orbital $\alpha$
at site $i$, and $\mathbf{e}_{i}$ is the prescribed moment direction.
We use $I_{d}=0.96\;\text{eV}$ for the $d$ orbitals and $I_{s}=I_{p}=I_{d}/10$
for the $s$ and $p$ orbitals \cite{Aute2006,Schena2010}. The spin operator reads
\begin{equation}
\hat{\mathbf{S}}_{i,\alpha}=\frac{\hbar}{2}\sum_{\sigma\sigma'}\boldsymbol{\sigma}_{\sigma\sigma'}\hat{c}_{i\alpha\sigma}^{\dagger}\hat{c}_{i\alpha\sigma'},
\end{equation}
with Pauli matrix vector $\boldsymbol{\sigma}$. Charge neutrality
is enforced by the term
\begin{equation}
\hat{\mathcal{H}}_{\text{lcn}}=U_{\text{lcn}}\sum_{i}\hat{n}_{i}\left(n_{i}-n_{i}^{0}\right),
\end{equation}
where 
\begin{equation}
\hat{n}_{i}=\sum_{\alpha,\sigma}\hat{c}_{i\alpha\sigma}^{\dagger}\hat{c}_{i\alpha\sigma}
\end{equation}
counts the number of electrons at site $i$ and $n_{i}^{0}$ is the
prescribed number of electrons per site. We use in our calculations
$U_{\text{lcn}}=5\;\text{eV}$ \cite{Aute2006,Schena2010}.

Both $\hat{\mathcal{H}}_{\text{Stoner}}$ and $\hat{\mathcal{H}}_{\text{lcn}}$ depend on the moment configuration $\{\mathbf{e}_i\}$, explicitly and via the charges $n_i$ and moment lengths $M_i$, which leads to a difference between $\mathbf{B}_{i}^{\text{grad}}$ and $-\mathbf{B}_{i}^{\text{con}}$ according to Eq.~(\ref{eq:constraining field theorem}).

\begin{figure}
\begin{centering}
\includegraphics[scale=1]{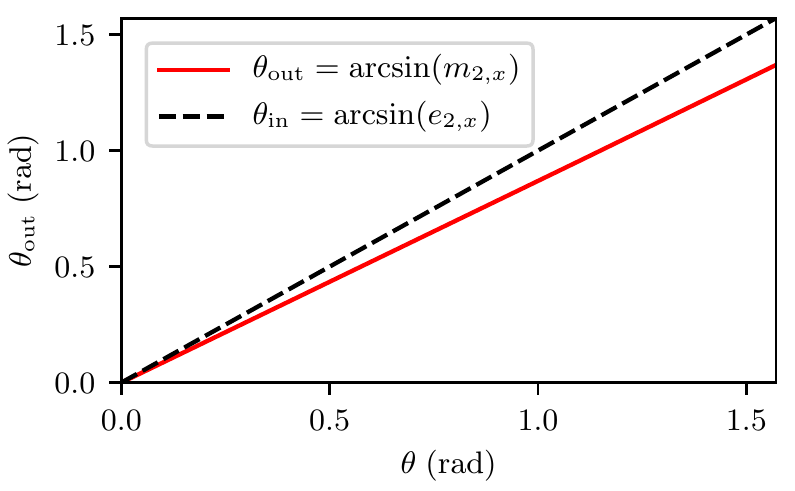}
\par\end{centering}
\caption{Comparison of the input and output moment directions $\mathbf{e}_2$ and $\mathbf{m}_2$, where the moment $i=2$ of an iron dimer is rotated by $\theta$, calculated with the tight-binding model without constraining fields. \label{fig:inout}}
\end{figure}

This tight-binding model already includes an approximate implementation
of the constrained moment directions $\left\{ \mathbf{e}_{i}\right\} $,
since these directions are favored by the Stoner term (\ref{eq:Stoner}).
Even without constraining fields, the moments align approximately
along those directions. This alignment is not exact, as shown in Fig.~\ref{fig:inout},
which implies that constraining fields are still required for accurate
results. Our implementation of the constraining fields for the tight-binding model is based on the method by Stocks \textit{et al.} \cite{Stocks1998,Ujfalussy1999}, as given by Eq.~(\ref{eq:Stocks}).

\section{Mean-field decoupling\label{sec:Mf decoupling}}

Here, we demonstrate that the term $\langle\boldsymbol{\nabla}_{\mathbf{e}_{i}}\hat{\mathcal{H}}_{\text{mf}}\rangle$  vanishes when the constant energy contributions arising from the mean-field decoupling are taken into account. We consider as an example the following mean-field decoupling of a Hubbard interaction term,
\begin{align}
    \hat{\mathcal{H}}_{U}	&=U\sum_{i}\hat{n}_{i\uparrow}\hat{n}_{i\downarrow}\nonumber\\
	&\approx U\sum_{i}\left(n_{i\uparrow}\hat{n}_{i\downarrow}+n_{i\downarrow}\hat{n}_{i\uparrow}-n_{i\uparrow}n_{i\downarrow}\right),\label{eq:Hubbard}
\end{align}
where the term $n_{i\uparrow}n_{i\downarrow}$ leads to the cancellation,
\begin{align}
   \left\langle\boldsymbol{\nabla}_{\mathbf{e}_{i}}\hat{\mathcal{H}}_U\right\rangle = U \sum_i \left[  n_{i\downarrow}\boldsymbol{\nabla}_{\mathbf{e}_{i}}n_{i\uparrow}+n_{i\uparrow}\boldsymbol{\nabla}_{\mathbf{e}_{i}}n_{i\downarrow}\right. \nonumber\\
   -\left. \boldsymbol{\nabla}_{\mathbf{e}_{i}}\left(n_{i\uparrow}n_{i\downarrow}\right)\right]=0.
\end{align}
The constant energy term in Eq.~(\ref{eq:Hubbard}) is required to avoid a double counting of energy contributions but does not change the dynamics of the Hamiltonian or the calculation of the constraining field.

\section{Numerical details\label{sec:Numerical_Details}}
Our first principles density functional theory computations are performed with the non-collinear implementation of the VASP program package \cite{Kresse1996,Kresse1996b}. The pseudopotential for Fe is considered in the generalized gradient approximation (GGA) of the projector augmented wave (PAW) method and with the cut-off energy of $600\;$eV. The valence electron configuration of the chosen Fe pseudopotential is $3d^7 4s^1$. For the dimer geometry we assume a distance of $2\;\textup{\AA}$ between the atoms, which are embedded in a cubic simulation box of $8\times 8 \times 8 \;\textup{\AA}^3$. Spin-orbit coupling is neglected in our setup. The constraining field implementation in VASP is based on Ref.~\cite{Ma2015} and the Lagrange multiplier $\lambda$ of this method is typically varied in the range $10-100\;\mathrm{eV}$. Figure \ref{fig:VASP} shows the results for $\lambda=50$ which are almost indistinguishable from the results for $\lambda=100$ indicating a sufficient convergence.

\section{Effective spin Hamiltonians\label{sec:Spin Hamiltonians}}

We define here an effective spin Hamiltonian $\mathcal{H}_{\text{eff}}$
as a classical spin model which provides the energy for a given moment
configuration $\left\{ \mathbf{e}_{i}\right\} $. The ideal effective
spin Hamiltonian would exactly reproduce the energies of the full
electronic Hamiltonian in the adiabatic approximation,
\begin{equation}
\mathcal{H}_{\text{eff}}(\left\{ \mathbf{e}_{i}\right\} )=\bra{\psi(\{\mathbf{e}_{i}\})}\hat{\mathcal{H}}_0\ket{\psi(\{\mathbf{e}_{i}\})}.
\end{equation}
Obtaining such an ideal Hamiltonian is of course in most cases impossible,
as it would require one to know the energy of each configuration $\left\{ \mathbf{e}_{i}\right\} $.
In practice, we have to rely on simple parametrizations of $\mathcal{H}_{\text{eff}}$
that capture the relevant behavior. 

For simplicity, we consider here only the Heisenberg model as an example,
\begin{equation}
\mathcal{H}_{\text{eff}}=-\frac{1}{2}\sum_{i,j}J_{ij}\mathbf{e}_{i}\cdot\mathbf{e}_{j},
\end{equation}
which is parameterized by the exchange constants $J_{ij}$.
The corresponding effective magnetic field is
\begin{align}
\mathbf{B}_{i}^{\text{eff}} &=-\frac{1}{M_{i}}\boldsymbol{\nabla}_{\mathbf{e}_{i}}\mathcal{H}_{\text{eff}}\nonumber\\
  &=\frac{1}{M_{i}}\sum_{j}J_{ij}\left(\mathbf{e}_{j}-\mathbf{e}_{i}(\mathbf{e}_{i}\cdot \mathbf{e}_{j})\right),
\end{align}
which requires knowledge of the magnetic moment length $M_{i}$. The
magnetic moment length itself may depend on the moment configuration,
$M_{i}=M_{i}(\left\{ \mathbf{e}_{i}\right\} )$, which makes this
approach only feasible if the moment length can be assumed to be constant. Here we had to subtract the component parallel to $\mathbf{e}_i$, which follows from the definition of the gradient (\ref{eq:gradient}) and the requirement that the effective field is perpendicular to $\mathbf{e}_i$.

For the dimer, the Heisenberg model depends only on a
single exchange constant $J$,
\begin{equation}
\mathcal{H}_{\text{eff}}=-J\mathbf{e}_{1}\cdot\mathbf{e}_{2},
\end{equation}
and the effective magnetic field is proportional to $J$,
\begin{align}
\mathbf{B}_{1}^{\text{eff}} &=\frac{J}{M_{1}}\left(\mathbf{e}_{2}-\mathbf{e}_{1}(\mathbf{e}_{1}\cdot\mathbf{e}_{2})\right),\\
\mathbf{B}_{2}^{\text{eff}} &=\frac{J}{M_{2}}\left(\mathbf{e}_{1}-\mathbf{e}_{2}(\mathbf{e}_{2}\cdot\mathbf{e}_{1})\right).
\end{align}


\begin{thebibliography}{35}%
\makeatletter
\providecommand \@ifxundefined [1]{%
 \@ifx{#1\undefined}
}%
\providecommand \@ifnum [1]{%
 \ifnum #1\expandafter \@firstoftwo
 \else \expandafter \@secondoftwo
 \fi
}%
\providecommand \@ifx [1]{%
 \ifx #1\expandafter \@firstoftwo
 \else \expandafter \@secondoftwo
 \fi
}%
\providecommand \natexlab [1]{#1}%
\providecommand \enquote  [1]{``#1''}%
\providecommand \bibnamefont  [1]{#1}%
\providecommand \bibfnamefont [1]{#1}%
\providecommand \citenamefont [1]{#1}%
\providecommand \href@noop [0]{\@secondoftwo}%
\providecommand \href [0]{\begingroup \@sanitize@url \@href}%
\providecommand \@href[1]{\@@startlink{#1}\@@href}%
\providecommand \@@href[1]{\endgroup#1\@@endlink}%
\providecommand \@sanitize@url [0]{\catcode `\\12\catcode `\$12\catcode
  `\&12\catcode `\#12\catcode `\^12\catcode `\_12\catcode `\%12\relax}%
\providecommand \@@startlink[1]{}%
\providecommand \@@endlink[0]{}%
\providecommand \url  [0]{\begingroup\@sanitize@url \@url }%
\providecommand \@url [1]{\endgroup\@href {#1}{\urlprefix }}%
\providecommand \urlprefix  [0]{URL }%
\providecommand \Eprint [0]{\href }%
\providecommand \doibase [0]{https://doi.org/}%
\providecommand \selectlanguage [0]{\@gobble}%
\providecommand \bibinfo  [0]{\@secondoftwo}%
\providecommand \bibfield  [0]{\@secondoftwo}%
\providecommand \translation [1]{[#1]}%
\providecommand \BibitemOpen [0]{}%
\providecommand \bibitemStop [0]{}%
\providecommand \bibitemNoStop [0]{.\EOS\space}%
\providecommand \EOS [0]{\spacefactor3000\relax}%
\providecommand \BibitemShut  [1]{\csname bibitem#1\endcsname}%
\let\auto@bib@innerbib\@empty
\bibitem [{\citenamefont {Bertotti}\ \emph {et~al.}(2009)\citenamefont
  {Bertotti}, \citenamefont {Mayergoyz},\ and\ \citenamefont
  {Serpico}}]{Bertotti2009}%
  \BibitemOpen
  \bibfield  {author} {\bibinfo {author} {\bibfnamefont {G.}~\bibnamefont
  {Bertotti}}, \bibinfo {author} {\bibfnamefont {I.~D.}\ \bibnamefont
  {Mayergoyz}},\ and\ \bibinfo {author} {\bibfnamefont {C.}~\bibnamefont
  {Serpico}},\ }\href {https://doi.org/10.1016/B978-0-08-044316-4.X0001-1}
  {\emph {\bibinfo {title} {Nonlinear Magnetization Dynamics in Nanosystems}}}\
  (\bibinfo  {publisher} {Elsevier},\ \bibinfo {address} {Oxford},\ \bibinfo
  {year} {2009})\BibitemShut {NoStop}%
\bibitem [{\citenamefont {Eriksson}\ \emph {et~al.}(2017)\citenamefont
  {Eriksson}, \citenamefont {Bergman}, \citenamefont {Bergqvist},\ and\
  \citenamefont {Hellsvik}}]{Eriksson2017}%
  \BibitemOpen
  \bibfield  {author} {\bibinfo {author} {\bibfnamefont {O.}~\bibnamefont
  {Eriksson}}, \bibinfo {author} {\bibfnamefont {A.}~\bibnamefont {Bergman}},
  \bibinfo {author} {\bibfnamefont {L.}~\bibnamefont {Bergqvist}},\ and\
  \bibinfo {author} {\bibfnamefont {J.}~\bibnamefont {Hellsvik}},\ }\href
  {https://doi.org/10.1093/oso/9780198788669.001.0001} {\emph {\bibinfo {title}
  {Atomistic Spin Dynamics: Foundations and Applications}}}\ (\bibinfo
  {publisher} {Oxford University Press, Oxford},\ \bibinfo {year} {2017})\BibitemShut
  {NoStop}%
\bibitem [{\citenamefont {Szilva}\ \emph {et~al.}(2013)\citenamefont {Szilva},
  \citenamefont {Costa}, \citenamefont {Bergman}, \citenamefont {Szunyogh},
  \citenamefont {Nordstr\"om},\ and\ \citenamefont {Eriksson}}]{Szilva2013}%
  \BibitemOpen
  \bibfield  {author} {\bibinfo {author} {\bibfnamefont {A.}~\bibnamefont
  {Szilva}}, \bibinfo {author} {\bibfnamefont {M.}~\bibnamefont {Costa}},
  \bibinfo {author} {\bibfnamefont {A.}~\bibnamefont {Bergman}}, \bibinfo
  {author} {\bibfnamefont {L.}~\bibnamefont {Szunyogh}}, \bibinfo {author}
  {\bibfnamefont {L.}~\bibnamefont {Nordstr\"om}},\ and\ \bibinfo {author}
  {\bibfnamefont {O.}~\bibnamefont {Eriksson}},\ }\bibfield  {title} {\bibinfo
  {title} {{Interatomic Exchange Interactions for Finite-Temperature Magnetism
  and Nonequilibrium Spin Dynamics}},\ }\href
  {https://doi.org/10.1103/PhysRevLett.111.127204} {\bibfield  {journal}
  {\bibinfo  {journal} {Phys. Rev. Lett.}\ }\textbf {\bibinfo {volume} {111}},\
  \bibinfo {pages} {127204} (\bibinfo {year} {2013})}\BibitemShut {NoStop}%
\bibitem [{\citenamefont {Szilva}\ \emph {et~al.}(2017)\citenamefont {Szilva},
  \citenamefont {Thonig}, \citenamefont {Bessarab}, \citenamefont {Kvashnin},
  \citenamefont {Rodrigues}, \citenamefont {Cardias}, \citenamefont {Pereiro},
  \citenamefont {Nordstr\"om}, \citenamefont {Bergman}, \citenamefont
  {Klautau},\ and\ \citenamefont {Eriksson}}]{Szilva2017}%
  \BibitemOpen
  \bibfield  {author} {\bibinfo {author} {\bibfnamefont {A.}~\bibnamefont
  {Szilva}}, \bibinfo {author} {\bibfnamefont {D.}~\bibnamefont {Thonig}},
  \bibinfo {author} {\bibfnamefont {P.~F.}\ \bibnamefont {Bessarab}}, \bibinfo
  {author} {\bibfnamefont {Y.~O.}\ \bibnamefont {Kvashnin}}, \bibinfo {author}
  {\bibfnamefont {D.~C.~M.}\ \bibnamefont {Rodrigues}}, \bibinfo {author}
  {\bibfnamefont {R.}~\bibnamefont {Cardias}}, \bibinfo {author} {\bibfnamefont
  {M.}~\bibnamefont {Pereiro}}, \bibinfo {author} {\bibfnamefont
  {L.}~\bibnamefont {Nordstr\"om}}, \bibinfo {author} {\bibfnamefont
  {A.}~\bibnamefont {Bergman}}, \bibinfo {author} {\bibfnamefont {A.~B.}\
  \bibnamefont {Klautau}},\ and\ \bibinfo {author} {\bibfnamefont
  {O.}~\bibnamefont {Eriksson}},\ }\bibfield  {title} {\bibinfo {title}
  {{Theory of noncollinear interactions beyond Heisenberg exchange:
  Applications to bcc Fe}},\ }\href
  {https://doi.org/10.1103/PhysRevB.96.144413} {\bibfield  {journal} {\bibinfo
  {journal} {Phys. Rev. B}\ }\textbf {\bibinfo {volume} {96}},\ \bibinfo
  {pages} {144413} (\bibinfo {year} {2017})}\BibitemShut {NoStop}%
\bibitem [{\citenamefont {Beaurepaire}\ \emph {et~al.}(1996)\citenamefont
  {Beaurepaire}, \citenamefont {Merle}, \citenamefont {Daunois},\ and\
  \citenamefont {Bigot}}]{Beaurepaire1996}%
  \BibitemOpen
  \bibfield  {author} {\bibinfo {author} {\bibfnamefont {E.}~\bibnamefont
  {Beaurepaire}}, \bibinfo {author} {\bibfnamefont {J.-C.}\ \bibnamefont
  {Merle}}, \bibinfo {author} {\bibfnamefont {A.}~\bibnamefont {Daunois}},\
  and\ \bibinfo {author} {\bibfnamefont {J.-Y.}\ \bibnamefont {Bigot}},\
  }\bibfield  {title} {\bibinfo {title} {Ultrafast spin dynamics in
  ferromagnetic nickel},\ }\href {https://doi.org/10.1103/PhysRevLett.76.4250}
  {\bibfield  {journal} {\bibinfo  {journal} {Phys. Rev. Lett.}\ }\textbf
  {\bibinfo {volume} {76}},\ \bibinfo {pages} {4250} (\bibinfo {year}
  {1996})}\BibitemShut {NoStop}%
\bibitem [{\citenamefont {Runge}\ and\ \citenamefont
  {Gross}(1984)}]{Runge1984}%
  \BibitemOpen
  \bibfield  {author} {\bibinfo {author} {\bibfnamefont {E.}~\bibnamefont
  {Runge}}\ and\ \bibinfo {author} {\bibfnamefont {E.~K.~U.}\ \bibnamefont
  {Gross}},\ }\bibfield  {title} {\bibinfo {title} {Density-functional theory
  for time-dependent systems},\ }\href
  {https://doi.org/10.1103/PhysRevLett.52.997} {\bibfield  {journal} {\bibinfo
  {journal} {Phys. Rev. Lett.}\ }\textbf {\bibinfo {volume} {52}},\ \bibinfo
  {pages} {997} (\bibinfo {year} {1984})}\BibitemShut {NoStop}%
\bibitem [{\citenamefont {Antropov}\ \emph {et~al.}(1995)\citenamefont
  {Antropov}, \citenamefont {Katsnelson}, \citenamefont {van Schilfgaarde},\
  and\ \citenamefont {Harmon}}]{Antropov1995}%
  \BibitemOpen
  \bibfield  {author} {\bibinfo {author} {\bibfnamefont {V.~P.}\ \bibnamefont
  {Antropov}}, \bibinfo {author} {\bibfnamefont {M.~I.}\ \bibnamefont
  {Katsnelson}}, \bibinfo {author} {\bibfnamefont {M.}~\bibnamefont {van
  Schilfgaarde}},\ and\ \bibinfo {author} {\bibfnamefont {B.~N.}\ \bibnamefont
  {Harmon}},\ }\bibfield  {title} {\bibinfo {title} {{$\mathit{Ab}$
  $\mathit{Initio}$ Spin Dynamics in Magnets}},\ }\href
  {https://doi.org/10.1103/PhysRevLett.75.729} {\bibfield  {journal} {\bibinfo
  {journal} {Phys. Rev. Lett.}\ }\textbf {\bibinfo {volume} {75}},\ \bibinfo
  {pages} {729} (\bibinfo {year} {1995})}\BibitemShut {NoStop}%
\bibitem [{\citenamefont {Antropov}\ \emph {et~al.}(1996)\citenamefont
  {Antropov}, \citenamefont {Katsnelson}, \citenamefont {Harmon}, \citenamefont
  {van Schilfgaarde},\ and\ \citenamefont {Kusnezov}}]{Antropov1996}%
  \BibitemOpen
  \bibfield  {author} {\bibinfo {author} {\bibfnamefont {V.~P.}\ \bibnamefont
  {Antropov}}, \bibinfo {author} {\bibfnamefont {M.~I.}\ \bibnamefont
  {Katsnelson}}, \bibinfo {author} {\bibfnamefont {B.~N.}\ \bibnamefont
  {Harmon}}, \bibinfo {author} {\bibfnamefont {M.}~\bibnamefont {van
  Schilfgaarde}},\ and\ \bibinfo {author} {\bibfnamefont {D.}~\bibnamefont
  {Kusnezov}},\ }\bibfield  {title} {\bibinfo {title} {Spin dynamics in
  magnets: Equation of motion and finite temperature effects},\ }\href
  {https://doi.org/10.1103/PhysRevB.54.1019} {\bibfield  {journal} {\bibinfo
  {journal} {Phys. Rev. B}\ }\textbf {\bibinfo {volume} {54}},\ \bibinfo
  {pages} {1019} (\bibinfo {year} {1996})}\BibitemShut {NoStop}%
\bibitem [{\citenamefont {Stocks}\ \emph {et~al.}(1998)\citenamefont {Stocks},
  \citenamefont {Ujfalussy}, \citenamefont {Wang}, \citenamefont {Nicholson},
  \citenamefont {Shelton}, \citenamefont {Wang}, \citenamefont {Canning},\ and\
  \citenamefont {Gy\"orffy}}]{Stocks1998}%
  \BibitemOpen
  \bibfield  {author} {\bibinfo {author} {\bibfnamefont {G.~M.}\ \bibnamefont
  {Stocks}}, \bibinfo {author} {\bibfnamefont {B.}~\bibnamefont {Ujfalussy}},
  \bibinfo {author} {\bibfnamefont {X.}~\bibnamefont {Wang}}, \bibinfo {author}
  {\bibfnamefont {D.~M.~C.}\ \bibnamefont {Nicholson}}, \bibinfo {author}
  {\bibfnamefont {W.~A.}\ \bibnamefont {Shelton}}, \bibinfo {author}
  {\bibfnamefont {Y.}~\bibnamefont {Wang}}, \bibinfo {author} {\bibfnamefont
  {A.}~\bibnamefont {Canning}},\ and\ \bibinfo {author} {\bibfnamefont {B.~L.}\
  \bibnamefont {Gy\"orffy}},\ }\bibfield  {title} {\bibinfo {title} {Towards a
  constrained local moment model for first principles spin dynamics},\ }\href
  {https://doi.org/10.1080/13642819808206775} {\bibfield  {journal} {\bibinfo
  {journal} {Philos. Mag. B}\ }\textbf {\bibinfo {volume} {78}},\ \bibinfo
  {pages} {665} (\bibinfo {year} {1998})}\BibitemShut {NoStop}%
\bibitem [{\citenamefont {Ujfalussy}\ \emph {et~al.}(1999)\citenamefont
  {Ujfalussy}, \citenamefont {Wang}, \citenamefont {Nicholson}, \citenamefont
  {Shelton}, \citenamefont {Stocks}, \citenamefont {Wang},\ and\ \citenamefont
  {Gyorffy}}]{Ujfalussy1999}%
  \BibitemOpen
  \bibfield  {author} {\bibinfo {author} {\bibfnamefont {B.}~\bibnamefont
  {Ujfalussy}}, \bibinfo {author} {\bibfnamefont {X.-D.}\ \bibnamefont {Wang}},
  \bibinfo {author} {\bibfnamefont {D.~M.~C.}\ \bibnamefont {Nicholson}},
  \bibinfo {author} {\bibfnamefont {W.~A.}\ \bibnamefont {Shelton}}, \bibinfo
  {author} {\bibfnamefont {G.~M.}\ \bibnamefont {Stocks}}, \bibinfo {author}
  {\bibfnamefont {Y.}~\bibnamefont {Wang}},\ and\ \bibinfo {author}
  {\bibfnamefont {B.~L.}\ \bibnamefont {Gyorffy}},\ }\bibfield  {title}
  {\bibinfo {title} {Constrained density functional theory for first principles
  spin dynamics},\ }\href {https://doi.org/10.1063/1.370494} {\bibfield
  {journal} {\bibinfo  {journal} {J. Appl. Phys.}\ }\textbf {\bibinfo {volume}
  {85}},\ \bibinfo {pages} {4824} (\bibinfo {year} {1999})}\BibitemShut
  {NoStop}%
\bibitem [{\citenamefont {Dederichs}\ \emph {et~al.}(1984)\citenamefont
  {Dederichs}, \citenamefont {Bl\"ugel}, \citenamefont {Zeller},\ and\
  \citenamefont {Akai}}]{Dederichs1984}%
  \BibitemOpen
  \bibfield  {author} {\bibinfo {author} {\bibfnamefont {P.~H.}\ \bibnamefont
  {Dederichs}}, \bibinfo {author} {\bibfnamefont {S.}~\bibnamefont {Bl\"ugel}},
  \bibinfo {author} {\bibfnamefont {R.}~\bibnamefont {Zeller}},\ and\ \bibinfo
  {author} {\bibfnamefont {H.}~\bibnamefont {Akai}},\ }\bibfield  {title}
  {\bibinfo {title} {Ground states of constrained systems: Application to
  cerium impurities},\ }\href {https://doi.org/10.1103/PhysRevLett.53.2512}
  {\bibfield  {journal} {\bibinfo  {journal} {Phys. Rev. Lett.}\ }\textbf
  {\bibinfo {volume} {53}},\ \bibinfo {pages} {2512} (\bibinfo {year}
  {1984})}\BibitemShut {NoStop}%
\bibitem [{\citenamefont {Singer}\ \emph {et~al.}(2005)\citenamefont {Singer},
  \citenamefont {F\"ahnle},\ and\ \citenamefont {Bihlmayer}}]{Singer2005}%
  \BibitemOpen
  \bibfield  {author} {\bibinfo {author} {\bibfnamefont {R.}~\bibnamefont
  {Singer}}, \bibinfo {author} {\bibfnamefont {M.}~\bibnamefont {F\"ahnle}},\
  and\ \bibinfo {author} {\bibfnamefont {G.}~\bibnamefont {Bihlmayer}},\
  }\bibfield  {title} {\bibinfo {title} {Constrained spin-density functional
  theory for excited magnetic configurations in an adiabatic approximation},\
  }\href {https://doi.org/10.1103/PhysRevB.71.214435} {\bibfield  {journal}
  {\bibinfo  {journal} {Phys. Rev. B}\ }\textbf {\bibinfo {volume} {71}},\
  \bibinfo {pages} {214435} (\bibinfo {year} {2005})}\BibitemShut {NoStop}%
\bibitem [{\citenamefont {Halilov}\ \emph {et~al.}(1998)\citenamefont
  {Halilov}, \citenamefont {Eschrig}, \citenamefont {Perlov},\ and\
  \citenamefont {Oppeneer}}]{Halilov1998}%
  \BibitemOpen
  \bibfield  {author} {\bibinfo {author} {\bibfnamefont {S.~V.}\ \bibnamefont
  {Halilov}}, \bibinfo {author} {\bibfnamefont {H.}~\bibnamefont {Eschrig}},
  \bibinfo {author} {\bibfnamefont {A.~Y.}\ \bibnamefont {Perlov}},\ and\
  \bibinfo {author} {\bibfnamefont {P.~M.}\ \bibnamefont {Oppeneer}},\
  }\bibfield  {title} {\bibinfo {title} {{Adiabatic spin dynamics from
  spin-density-functional theory: Application to Fe, Co, and Ni}},\ }\href
  {https://doi.org/10.1103/PhysRevB.58.293} {\bibfield  {journal} {\bibinfo
  {journal} {Phys. Rev. B}\ }\textbf {\bibinfo {volume} {58}},\ \bibinfo
  {pages} {293} (\bibinfo {year} {1998})}\BibitemShut {NoStop}%
\bibitem [{\citenamefont {Landau}\ and\ \citenamefont
  {Lifshitz}(1935)}]{Landau35}%
  \BibitemOpen
  \bibfield  {author} {\bibinfo {author} {\bibfnamefont {L.~D.}\ \bibnamefont
  {Landau}}\ and\ \bibinfo {author} {\bibfnamefont {E.}~\bibnamefont
  {Lifshitz}},\ }\bibfield  {title} {\bibinfo {title} {On the theory of the
  dispersion of magnetic permeability in ferromagnetic bodies},\ }\href
  {http://archive.ujp.bitp.kiev.ua/files/journals/53/si/53SI06p.pdf} {\bibfield
   {journal} {\bibinfo  {journal} {Phys. Z. Sowjet.}\ }\textbf {\bibinfo
  {volume} {8}},\ \bibinfo {pages} {153} (\bibinfo {year} {1935})}\BibitemShut
  {NoStop}%
\bibitem [{\citenamefont {Keshtgar}\ \emph {et~al.}(2017)\citenamefont
  {Keshtgar}, \citenamefont {Streib}, \citenamefont {Kamra}, \citenamefont
  {Blanter},\ and\ \citenamefont {Bauer}}]{Keshtgar2017}%
  \BibitemOpen
  \bibfield  {author} {\bibinfo {author} {\bibfnamefont {H.}~\bibnamefont
  {Keshtgar}}, \bibinfo {author} {\bibfnamefont {S.}~\bibnamefont {Streib}},
  \bibinfo {author} {\bibfnamefont {A.}~\bibnamefont {Kamra}}, \bibinfo
  {author} {\bibfnamefont {Y.~M.}\ \bibnamefont {Blanter}},\ and\ \bibinfo
  {author} {\bibfnamefont {G.~E.~W.}\ \bibnamefont {Bauer}},\ }\bibfield
  {title} {\bibinfo {title} {Magnetomechanical coupling and ferromagnetic
  resonance in magnetic nanoparticles},\ }\href
  {https://doi.org/10.1103/PhysRevB.95.134447} {\bibfield  {journal} {\bibinfo
  {journal} {Phys. Rev. B}\ }\textbf {\bibinfo {volume} {95}},\ \bibinfo
  {pages} {134447} (\bibinfo {year} {2017})}\BibitemShut {NoStop}%
\bibitem [{\citenamefont {Ma}\ and\ \citenamefont {Dudarev}(2015)}]{Ma2015}%
  \BibitemOpen
  \bibfield  {author} {\bibinfo {author} {\bibfnamefont {P.-W.}\ \bibnamefont
  {Ma}}\ and\ \bibinfo {author} {\bibfnamefont {S.~L.}\ \bibnamefont
  {Dudarev}},\ }\bibfield  {title} {\bibinfo {title} {Constrained density
  functional for noncollinear magnetism},\ }\href
  {https://doi.org/10.1103/PhysRevB.91.054420} {\bibfield  {journal} {\bibinfo
  {journal} {Phys. Rev. B}\ }\textbf {\bibinfo {volume} {91}},\ \bibinfo
  {pages} {054420} (\bibinfo {year} {2015})}\BibitemShut {NoStop}%
\bibitem [{\citenamefont {Ivanov}\ \emph {et~al.}(2020)\citenamefont {Ivanov},
  \citenamefont {Bessarab.}, \citenamefont {J\'onsson},\ and\ \citenamefont
  {Uzdin}}]{Ivanov2020}%
  \BibitemOpen
  \bibfield  {author} {\bibinfo {author} {\bibfnamefont {A.~V.}\ \bibnamefont
  {Ivanov}}, \bibinfo {author} {\bibfnamefont {P.~F.}\ \bibnamefont
  {Bessarab.}}, \bibinfo {author} {\bibfnamefont {H.}~\bibnamefont
  {J\'onsson}},\ and\ \bibinfo {author} {\bibfnamefont {V.~M.}\ \bibnamefont
  {Uzdin}},\ }\bibfield  {title} {\bibinfo {title} {{Fully self-consistent
  calculations of magnetic structure within non-collinear Alexander-Anderson
  model}},\ }\href {https://doi.org/10.17586/2220-8054-2020-11-1-65-77}
  {\bibfield  {journal} {\bibinfo  {journal} {Nanosyst. Phys. Chem. Math.}\
  }\textbf {\bibinfo {volume} {1}},\ \bibinfo {pages} {65} (\bibinfo {year}
  {2020})}\BibitemShut {NoStop}%
\bibitem [{\citenamefont {Wu}\ and\ \citenamefont
  {Van~Voorhis}(2005)}]{Wu2005}%
  \BibitemOpen
  \bibfield  {author} {\bibinfo {author} {\bibfnamefont {Q.}~\bibnamefont
  {Wu}}\ and\ \bibinfo {author} {\bibfnamefont {T.}~\bibnamefont
  {Van~Voorhis}},\ }\bibfield  {title} {\bibinfo {title} {Direct optimization
  method to study constrained systems within density-functional theory},\
  }\href {https://doi.org/10.1103/PhysRevA.72.024502} {\bibfield  {journal}
  {\bibinfo  {journal} {Phys. Rev. A}\ }\textbf {\bibinfo {volume} {72}},\
  \bibinfo {pages} {024502} (\bibinfo {year} {2005})}\BibitemShut {NoStop}%
\bibitem [{\citenamefont {Kurz}\ \emph {et~al.}(2004)\citenamefont {Kurz},
  \citenamefont {F\"orster}, \citenamefont {Nordstr\"om}, \citenamefont
  {Bihlmayer},\ and\ \citenamefont {Bl\"ugel}}]{Kurz2004}%
  \BibitemOpen
  \bibfield  {author} {\bibinfo {author} {\bibfnamefont {P.}~\bibnamefont
  {Kurz}}, \bibinfo {author} {\bibfnamefont {F.}~\bibnamefont {F\"orster}},
  \bibinfo {author} {\bibfnamefont {L.}~\bibnamefont {Nordstr\"om}}, \bibinfo
  {author} {\bibfnamefont {G.}~\bibnamefont {Bihlmayer}},\ and\ \bibinfo
  {author} {\bibfnamefont {S.}~\bibnamefont {Bl\"ugel}},\ }\bibfield  {title}
  {\bibinfo {title} {Ab initio treatment of noncollinear magnets with the
  full-potential linearized augmented plane wave method},\ }\href
  {https://doi.org/10.1103/PhysRevB.69.024415} {\bibfield  {journal} {\bibinfo
  {journal} {Phys. Rev. B}\ }\textbf {\bibinfo {volume} {69}},\ \bibinfo
  {pages} {024415} (\bibinfo {year} {2004})}\BibitemShut {NoStop}%
\bibitem [{\citenamefont {Cuadrado}\ \emph {et~al.}(2018)\citenamefont
  {Cuadrado}, \citenamefont {Pruneda}, \citenamefont {Garc{\'{\i}}a},\ and\
  \citenamefont {Ordej{\'{o}}n}}]{Cuadrado2018}%
  \BibitemOpen
  \bibfield  {author} {\bibinfo {author} {\bibfnamefont {R.}~\bibnamefont
  {Cuadrado}}, \bibinfo {author} {\bibfnamefont {M.}~\bibnamefont {Pruneda}},
  \bibinfo {author} {\bibfnamefont {A.}~\bibnamefont {Garc{\'{\i}}a}},\ and\
  \bibinfo {author} {\bibfnamefont {P.}~\bibnamefont {Ordej{\'{o}}n}},\
  }\bibfield  {title} {\bibinfo {title} {{Implementation of non-collinear
  spin-constrained {DFT} calculations in {SIESTA} with a fully relativistic
  Hamiltonian}},\ }\href {https://doi.org/10.1088/2515-7639/aae7db} {\bibfield
  {journal} {\bibinfo  {journal} {J. Phys. Mater.}\ }\textbf {\bibinfo {volume}
  {1}},\ \bibinfo {pages} {015010} (\bibinfo {year} {2018})}\BibitemShut
  {NoStop}%
\bibitem [{\citenamefont {Kresse}\ and\ \citenamefont
  {Furthm\"uller}(1996)}]{Kresse1996}%
  \BibitemOpen
  \bibfield  {author} {\bibinfo {author} {\bibfnamefont {G.}~\bibnamefont
  {Kresse}}\ and\ \bibinfo {author} {\bibfnamefont {J.}~\bibnamefont
  {Furthm\"uller}},\ }\bibfield  {title} {\bibinfo {title} {Efficient iterative
  schemes for ab initio total-energy calculations using a plane-wave basis
  set},\ }\href {https://doi.org/10.1103/PhysRevB.54.11169} {\bibfield
  {journal} {\bibinfo  {journal} {Phys. Rev. B}\ }\textbf {\bibinfo {volume}
  {54}},\ \bibinfo {pages} {11169} (\bibinfo {year} {1996})}\BibitemShut
  {NoStop}%
\bibitem [{\citenamefont {Kresse}\ and\ \citenamefont
  {Furthmüller}(1996)}]{Kresse1996b}%
  \BibitemOpen
  \bibfield  {author} {\bibinfo {author} {\bibfnamefont {G.}~\bibnamefont
  {Kresse}}\ and\ \bibinfo {author} {\bibfnamefont {J.}~\bibnamefont
  {Furthmüller}},\ }\bibfield  {title} {\bibinfo {title} {Efficiency of
  \textit{ab initio} total energy calculations for metals and semiconductors using a
  plane-wave basis set},\ }\href
  {https://doi.org/https://doi.org/10.1016/0927-0256(96)00008-0} {\bibfield
  {journal} {\bibinfo  {journal} {Comp. Mater. Sci.}\ }\textbf {\bibinfo
  {volume} {6}},\ \bibinfo {pages} {15 } (\bibinfo {year} {1996})}\BibitemShut
  {NoStop}%
\bibitem [{\citenamefont {Grotheer}\ \emph {et~al.}(2001)\citenamefont
  {Grotheer}, \citenamefont {Ederer},\ and\ \citenamefont
  {F\"ahnle}}]{Grotheer2001}%
  \BibitemOpen
  \bibfield  {author} {\bibinfo {author} {\bibfnamefont {O.}~\bibnamefont
  {Grotheer}}, \bibinfo {author} {\bibfnamefont {C.}~\bibnamefont {Ederer}},\
  and\ \bibinfo {author} {\bibfnamefont {M.}~\bibnamefont {F\"ahnle}},\
  }\bibfield  {title} {\bibinfo {title} {Fast ab initio methods for the
  calculation of adiabatic spin wave spectra in complex systems},\ }\href
  {https://doi.org/10.1103/PhysRevB.63.100401} {\bibfield  {journal} {\bibinfo
  {journal} {Phys. Rev. B}\ }\textbf {\bibinfo {volume} {63}},\ \bibinfo
  {pages} {100401} (\bibinfo {year} {2001})}\BibitemShut {NoStop}%
\bibitem [{\citenamefont {Liechtenstein}\ \emph {et~al.}(1984)\citenamefont
  {Liechtenstein}, \citenamefont {Katsnelson},\ and\ \citenamefont
  {Gubanov}}]{Liechtenstein1984}%
  \BibitemOpen
  \bibfield  {author} {\bibinfo {author} {\bibfnamefont {A.~I.}\ \bibnamefont
  {Liechtenstein}}, \bibinfo {author} {\bibfnamefont {M.~I.}\ \bibnamefont
  {Katsnelson}},\ and\ \bibinfo {author} {\bibfnamefont {V.~A.}\ \bibnamefont
  {Gubanov}},\ }\bibfield  {title} {\bibinfo {title} {Exchange interactions and
  spin-wave stiffness in ferromagnetic metals},\ }\href
  {https://doi.org/10.1088/0305-4608/14/7/007} {\bibfield  {journal} {\bibinfo
  {journal} {J. Phys. F}\ }\textbf {\bibinfo {volume} {14}},\ \bibinfo {pages}
  {L125} (\bibinfo {year} {1984})}\BibitemShut {NoStop}%
\bibitem [{\citenamefont {Liechtenstein}\ \emph {et~al.}(1987)\citenamefont
  {Liechtenstein}, \citenamefont {Katsnelson}, \citenamefont {Antropov},\ and\
  \citenamefont {Gubanov}}]{Liechtenstein1987}%
  \BibitemOpen
  \bibfield  {author} {\bibinfo {author} {\bibfnamefont {A.}~\bibnamefont
  {Liechtenstein}}, \bibinfo {author} {\bibfnamefont {M.}~\bibnamefont
  {Katsnelson}}, \bibinfo {author} {\bibfnamefont {V.}~\bibnamefont
  {Antropov}},\ and\ \bibinfo {author} {\bibfnamefont {V.}~\bibnamefont
  {Gubanov}},\ }\bibfield  {title} {\bibinfo {title} {Local spin density
  functional approach to the theory of exchange interactions in ferromagnetic
  metals and alloys},\ }\href
  {https://doi.org/https://doi.org/10.1016/0304-8853(87)90721-9} {\bibfield
  {journal} {\bibinfo  {journal} {J. Magn. Magn. Mater.}\ }\textbf {\bibinfo
  {volume} {67}},\ \bibinfo {pages} {65 } (\bibinfo {year} {1987})}\BibitemShut
  {NoStop}%
\bibitem [{\citenamefont {Bruno}(2003)}]{Bruno2003}%
  \BibitemOpen
  \bibfield  {author} {\bibinfo {author} {\bibfnamefont {P.}~\bibnamefont
  {Bruno}},\ }\bibfield  {title} {\bibinfo {title} {Exchange interaction
  parameters and adiabatic spin-wave spectra of ferromagnets: A ``renormalized
  magnetic force theorem''},\ }\href
  {https://doi.org/10.1103/PhysRevLett.90.087205} {\bibfield  {journal}
  {\bibinfo  {journal} {Phys. Rev. Lett.}\ }\textbf {\bibinfo {volume} {90}},\
  \bibinfo {pages} {087205} (\bibinfo {year} {2003})}\BibitemShut {NoStop}%
\bibitem [{\citenamefont {Jacobsson}\ \emph {et~al.}(2017)\citenamefont
  {Jacobsson}, \citenamefont {Johansson}, \citenamefont {Gorbatov},
  \citenamefont {Ležaić}, \citenamefont {Sanyal}, \citenamefont {Blügel},\
  and\ \citenamefont {Etz}}]{Jacobsson2017}%
  \BibitemOpen
  \bibfield  {author} {\bibinfo {author} {\bibfnamefont {A.}~\bibnamefont
  {Jacobsson}}, \bibinfo {author} {\bibfnamefont {G.}~\bibnamefont
  {Johansson}}, \bibinfo {author} {\bibfnamefont {O.~I.}\ \bibnamefont
  {Gorbatov}}, \bibinfo {author} {\bibfnamefont {M.}~\bibnamefont {Ležaić}},
  \bibinfo {author} {\bibfnamefont {B.}~\bibnamefont {Sanyal}}, \bibinfo
  {author} {\bibfnamefont {S.}~\bibnamefont {Blügel}},\ and\ \bibinfo {author}
  {\bibfnamefont {C.}~\bibnamefont {Etz}},\ }\href@noop {} {\bibinfo {title}
  {Parameterisation of non-collinear energy landscapes in itinerant magnets}}
  (\bibinfo {year} {2017}),\ \Eprint {https://arxiv.org/abs/arXiv:1702.00599}
  {arXiv:1702.00599} \BibitemShut {NoStop}%
\bibitem [{\citenamefont {Hellmann}(1937)}]{Hellmann1937}%
  \BibitemOpen
  \bibfield  {author} {\bibinfo {author} {\bibfnamefont {H.}~\bibnamefont
  {Hellmann}},\ }\href@noop {} {\emph {\bibinfo {title} {Einf\"uhrung in die
  Quantenchemie}}}\ [English translation: Introduction to Quantum Chemistry] (\bibinfo  {publisher} {Deuticke},\ \bibinfo {address}
  {Leipzig},\ \bibinfo {year} {1937})\BibitemShut {NoStop}%
\bibitem [{\citenamefont {Feynman}(1939)}]{Feynman1939}%
  \BibitemOpen
  \bibfield  {author} {\bibinfo {author} {\bibfnamefont {R.~P.}\ \bibnamefont
  {Feynman}},\ }\bibfield  {title} {\bibinfo {title} {Forces in molecules},\
  }\href {https://doi.org/10.1103/PhysRev.56.340} {\bibfield  {journal}
  {\bibinfo  {journal} {Phys. Rev.}\ }\textbf {\bibinfo {volume} {56}},\
  \bibinfo {pages} {340} (\bibinfo {year} {1939})}\BibitemShut {NoStop}%
\bibitem [{\citenamefont {Kohn}\ and\ \citenamefont {Sham}(1965)}]{Kohn1965}%
  \BibitemOpen
  \bibfield  {author} {\bibinfo {author} {\bibfnamefont {W.}~\bibnamefont
  {Kohn}}\ and\ \bibinfo {author} {\bibfnamefont {L.~J.}\ \bibnamefont
  {Sham}},\ }\bibfield  {title} {\bibinfo {title} {Self-consistent equations
  including exchange and correlation effects},\ }\href
  {https://doi.org/10.1103/PhysRev.140.A1133} {\bibfield  {journal} {\bibinfo
  {journal} {Phys. Rev.}\ }\textbf {\bibinfo {volume} {140}},\ \bibinfo {pages}
  {A1133} (\bibinfo {year} {1965})}\BibitemShut {NoStop}%
\bibitem [{Note1()}]{Note1}%
  \BibitemOpen
  \bibinfo {note} {The result $\langle \protect \boldsymbol {\nabla }_{\protect
  \mathbf {e}_{i}}\protect \hat {\protect \mathcal {H}}_{\protect \mathrm
  {KS}}\rangle +\protect \boldsymbol {\nabla }_{\protect \mathbf
  {e}_{i}}E_\protect \mathrm {dc}=\langle \protect \boldsymbol {\nabla
  }^*_{\protect \mathbf {e}_{i}}\protect \hat {\protect \mathcal {H}}_{\protect
  \mathrm {KS}}\rangle $ follows from the derivation in Appendix A of
  Ref.~\cite {Liechtenstein1987}.}\BibitemShut {Stop}%
\bibitem [{\citenamefont {Slater}\ and\ \citenamefont
  {Koster}(1954)}]{Slater1954}%
  \BibitemOpen
  \bibfield  {author} {\bibinfo {author} {\bibfnamefont {J.~C.}\ \bibnamefont
  {Slater}}\ and\ \bibinfo {author} {\bibfnamefont {G.~F.}\ \bibnamefont
  {Koster}},\ }\bibfield  {title} {\bibinfo {title} {{Simplified LCAO Method
  for the Periodic Potential Problem}},\ }\href
  {https://doi.org/10.1103/PhysRev.94.1498} {\bibfield  {journal} {\bibinfo
  {journal} {Phys. Rev.}\ }\textbf {\bibinfo {volume} {94}},\ \bibinfo {pages}
  {1498} (\bibinfo {year} {1954})}\BibitemShut {NoStop}%
\bibitem [{\citenamefont {Thonig}\ and\ \citenamefont
  {Henk}(2014)}]{Thonig2014}%
  \BibitemOpen
  \bibfield  {author} {\bibinfo {author} {\bibfnamefont {D.}~\bibnamefont
  {Thonig}}\ and\ \bibinfo {author} {\bibfnamefont {J.}~\bibnamefont {Henk}},\
  }\bibfield  {title} {\bibinfo {title} {{Gilbert damping tensor within the
  breathing Fermi surface model: anisotropy and non-locality}},\ }\href
  {https://doi.org/10.1088/1367-2630/16/1/013032} {\bibfield  {journal}
  {\bibinfo  {journal} {New J. Phys}\ }\textbf {\bibinfo {volume} {16}},\
  \bibinfo {pages} {013032} (\bibinfo {year} {2014})}\BibitemShut {NoStop}%
\bibitem [{\citenamefont {Aut{\`{e}}s}\ \emph {et~al.}(2006)\citenamefont
  {Aut{\`{e}}s}, \citenamefont {Barreteau}, \citenamefont {Spanjaard},\ and\
  \citenamefont {Desjonqu{\`{e}}res}}]{Aute2006}%
  \BibitemOpen
  \bibfield  {author} {\bibinfo {author} {\bibfnamefont {G.}~\bibnamefont
  {Aut{\`{e}}s}}, \bibinfo {author} {\bibfnamefont {C.}~\bibnamefont
  {Barreteau}}, \bibinfo {author} {\bibfnamefont {D.}~\bibnamefont
  {Spanjaard}},\ and\ \bibinfo {author} {\bibfnamefont {M.-C.}\ \bibnamefont
  {Desjonqu{\`{e}}res}},\ }\bibfield  {title} {\bibinfo {title} {Magnetism of
  iron: from the bulk to the monatomic wire},\ }\href
  {https://doi.org/10.1088/0953-8984/18/29/018} {\bibfield  {journal} {\bibinfo
   {journal} {J. Phys. Condens. Matter}\ }\textbf {\bibinfo {volume} {18}},\
  \bibinfo {pages} {6785} (\bibinfo {year} {2006})}\BibitemShut {NoStop}%
\bibitem [{\citenamefont {Schena}(2010)}]{Schena2010}%
  \BibitemOpen
  \bibfield  {author} {\bibinfo {author} {\bibfnamefont {T.}~\bibnamefont
  {Schena}},\ }\emph {\bibinfo {title} {{Tight-Binding Treatment of Complex
  Magnetic Structures in Low-Dimensional Systems}}},\ \href
  {https://www.fz-juelich.de/SharedDocs/Downloads/PGI/PGI-1/EN/Schena_diploma_pdf.pdf}
  {\bibinfo {type} {Diploma thesis}},\ \bibinfo  {school} {TH Aachen} (\bibinfo
  {year} {2010})\BibitemShut {NoStop}%
\end{thebibliography}
\end{document}